\documentclass[12pt,draftclsnofoot,journal,onecolumn]{IEEEtran}

\usepackage[dvips]{color}
\usepackage{epsf}
\usepackage{times}
\usepackage{epsfig}
\usepackage{graphicx}
\usepackage{url}
\usepackage{amsmath}
\usepackage{amssymb}
\usepackage{amsxtra}
\usepackage{algorithmic}
\usepackage{multirow}
\usepackage{cite}
\usepackage{enumerate}
\usepackage{subfigure}
\usepackage[ruled]{algorithm}

\newtheorem{definition}{\bf Definition}
\newtheorem{remark}{\bf Remark}

\renewcommand{\baselinestretch}{1.3} 

\newlength{\aligntop}
\newlength{\alignbot}
\makeatletter
\renewenvironment{align}{%
  \vspace{\aligntop}
  \start@align\@ne\st@rredfalse\m@ne
}{%
  \math@cr \black@\totwidth@
  \egroup
  \ifingather@
    \restorealignstate@
    \egroup
    \nonumber
    \ifnum0=`{\fi\iffalse}\fi
  \else
    $$%
  \fi
  \ignorespacesafterend%
  \vspace{\alignbot}\par\noindent
}

\IEEEoverridecommandlockouts
\begin{document}

\title{Spectrum Leasing as an Incentive towards Uplink Macrocell and Femtocell Cooperation}
\author{\authorblockN{Francesco Pantisano$^{1,2}$, Mehdi Bennis$^{1}$, Walid Saad$^\textbf{3}$ and M{\'e}rouane Debbah$^\textbf{4}$\\} \authorblockA{\small
$^\textbf{1}$ Centre for Wireless Communications - CWC, University of Oulu, 90570, Finland, email: \url{{fpantisa,bennis}@ee.oulu.fi}\\
$^\textbf{2}$ Dipartimento di Ingegneria dell'Energia Elettrica e dell'Informazione - DEI, University of Bologna, 40135, Italy, email:\url{francesco.pantisano@unibo.it}\\
$^\textbf{3}$ Electrical and Computer Engineering Department, University of Miami, Coral Gables, FL, USA, email:\url{walid@miami.edu}.\\
$^\textbf{4}$ Alcatel-Lucent Chair in Flexible Radio,  SUP{\'E}LEC, Gif-sur-Yvette, France email: \url{merouane.debbah@supelec.fr  }
 }%
   \thanks{ The authors would like to thank the Finnish funding agency for technology and innovation, Elektrobit, Nokia and Nokia Siemens Networks for supporting this work. This work has been performed in the framework of the ICT project
ICT-4-248523 BeFEMTO, which is partly funded by the EU.}}
\date{}
\maketitle

\thispagestyle{empty}

\begin{abstract}
The concept of femtocell access points underlaying existing communication infrastructure has recently emerged as a key technology that can significantly improve the coverage and performance of next-generation wireless networks. In this paper, we propose a framework for macrocell-femtocell cooperation under a closed access policy, in which a femtocell user may act as a relay for macrocell users. In return, each cooperative macrocell user grants the femtocell user a fraction of its superframe. We formulate a coalitional game with macrocell and femtocell users being the players, which can take individual and distributed decisions on whether to cooperate or not, while maximizing a utility function that captures the cooperative gains, in terms of throughput and delay. We show that the network can self-organize into a partition composed of disjoint coalitions which constitutes the \emph{recursive core} of the game representing a key solution concept for coalition formation games in partition form. Simulation results show that the proposed coalition formation algorithm yields significant gains in terms of average rate per macrocell user, reaching up to $239\%$, relative to the non-cooperative case. Moreover, the proposed approach shows an improvement in terms of femtocell users' rate of up to $21\%$ when compared to the traditional closed access policy.
\end{abstract}

\indent \indent 
 {\bf \small Index terms:} {\small spectrum leasing; femtocell networks; coalitional game theory; Device-to-Device~(D2D); cooperation; recursive core.}

\newpage
 \setcounter{page}{1}
\section{Introduction}
The new shifts in wireless communication paradigms, the need for energy-efficient communication, and the ever-increasing demands for ubiquitous wireless access and higher data rates led to increased research tackling the problem of deploying low cost, low power, femtocells. In fact, the challenges of introducing femtocell access points have attracted an increased attention from both academia as well as standardization bodies such as the Third Generation Partnership Program~(3GPP). It is envisioned that by introducing small cells, i.e., femtocells, serviced by dedicated femtocell access points, high quality indoor coverage can be achieved without any extra investments in network infrastructure, such as adding base stations or deploying advanced antenna systems. In addition, due to their ability to connect to existing backhaul networks (e.g., DSL), femtocells are an enabler for offloading traffic from existing wireless systems (e.g., cellular networks) and, subsequently, they can improve both spectrum efficiency and network capacity. As a result, two-tier femtocell networks, which consist of a macrocell network underlaid on femtocell access points~(FAPs) are expected to lie at the heart of emerging wireless systems~\cite{industry,industry2,LTE-A2,zhangg,befemto}.



The deployment of femtocells underlaid with an existing macrocell wireless network is accompanied with numerous technical challenges at different levels, such as spectrum allocation, handover, interference management and access policy. From a cross-tier interference standpoint, orthogonal spectrum allocation can entirely eliminate interference but is inefficient in terms of spectrum utilization \cite{Guvenc}. As a result, from a network operator's perspective, deploying co-channel femtocells is of great interest \cite{JA1,HC1}. Through co-channel operations, femtocell access points are able to reuse the spectral resources of the macrocell network and, hence, improve the spectral efficiency, especially when traffic loads are high. However, in co-channel operations, cross-tier (i.e., macro- to femtocell, and vice versa) as well as intra-tier (i.e., inter-femtocell) interference can seriously degrade the system performance \cite{choi,lp}. In order to mitigate both types of interference, a variety of decentralized solutions are provided in~\cite{sundeep,reed,garcia,dong,FP1,FP2,yangg,yook2,gustavo } and can be divided in two categories: self-organization and cooperative strategies. The former class includes non-cooperative mechanisms of adaptation to the operating scenario and the interference environment, and deals with dynamic spectrum occupation \cite{DSA} , power control \cite{yook2, Andrews2} and interference cancelation \cite{chan}. Self-organization relies on context awareness capabilities and the paradigm of learning  through trial-and-errors ~\cite{mehditao,gustavo,yangg}. Clearly, its key benefit is the scalability and the timeliness of the solutions, since the intelligence lies at the lower levels of the network architecture. One alternative approach is to leverage off the cooperation for interference management as done in \cite{FP2,Mischa,DP1}. Nevertheless, both self-organizing and cooperative solutions are associated with a cost or effort which can limit their benefits and, therefore, obtaining an optimal approach is quite challenging. In two-tier femtocell networks, different limitations can be witnessed at the mobile users' side. On one hand, macro users~(MUEs) are generally bandwidth limited and suffer from low signal to noise and interference ratios~(SINRs), especially when located at the cell boundary area. This is often reflected by a large number of outage events and a consequent increase in the user-plane delay. On the other hand, femtocell access points~(FAPs) are interference limited, therefore, a cross-tier cooperative scheme has to provide a benefit from different angles. In essence, some of the main open issues faced when designing cooperative schemes for femtocell networks are:

\begin{itemize}
  \item How can the cooperation among users belonging to different tiers be modeled?
  \item What is the price for cooperation and when is cooperation beneficial?
  \item How to provide incentives to encourage cooperation?
\end{itemize}


In femtocell deployment, three main access control policies for femtocells can be recognized \cite{dlr}: closed, open, and hybrid. In the closed access policy, femtocell subscribers constitute the closed subscribers group~(CSG) and are the only ones allowed to connect to the belonging femtocell. In the open access policy, non-subscribers may also connect to any femtocell, without any restriction. Lastly, in hybrid access policies, non-subscribers may connect to a femtocell only under particular circumstances, depending on the resource availability. One promising approach for cross-tier interference management is to enable open or hybrid access policies at the femtocells, so that the effects of macro- to femtocell interference is reduced as shown in~\cite{Andrews1,ping}, for open access. A general introduction on the issues of coexistence between macrocells and femtocells is provided in \cite{DP1,JA1}, in which the authors present various practical scenarios. Alternatively, the MBS can coordinate or direct main operations at the FAP by means of information exchange over the X2 interface or through the femtocell gateway \cite{ZGUO}. However, in large networks, the computational effort resulting from this procedure at the MBS can be high as it requires excessive traffic on the control channel. Hence, there is a need to develop cooperative strategies at the FAP level as it has been proposed recently in~\cite{JINJIN,LEE,bennis1}. A novel form of distributed compress-and forward scheme with decoder side information is studied in \cite{som} while further mechanisms of cooperation have been studied in the context of providing a reliable backhaul to the FAPs such as in~\cite{Osvaldo2}.




 The main contribution of this paper is to propose, within the context of wireless femtocell networks, a framework for macrocell-femtocell cooperation which allows to alleviate the uplink interference at the FAP and reduce the transmission delay at the macrouser. Unlike existing network architectures, we propose a model in which macro cellular users are granted femtocell access using a device-to-device~(D2D) link \cite{d2d} that enables them to communicate with a femtocell user~(FUE) that, in a second phase, acts as a relay for macrocell traffic. In essence, whenever an MUE and an FUE cooperate, the MUE forwards its own traffic to the FUE which, in turn, combines the MUE`s traffic with its own data and relays it to its serving FAP. This proposed concept allows the MUEs to explore nearby femtocells by cooperating with the FUEs, even when the FAPs adopt a closed-access policy and have a limited coverage area. Clearly, this scheme is beneficial for any MUE, located at the cell boundary area, that is suffering a low performance at its serving MBS and which is unable to connect to nearby FAP due to closed-access policy or limited FAP coverage. Therefore, the rationale behind the proposed approach is to capture an important mutual benefit in co-channel femtocell networks. On the one hand, in such underlay femtocell network, the availability of the spectral resources depends on the utilization of the macrocell tier, which is performed without cross-tier coordination. As a result, the number of users (both FUEs and guest MUEs) is ultimately limited by the FAP is ultimately limited by the interference produced by the MUEs. On the other hand, the performance of MUEs located at a cell edge is essentially limited by the achievable SINR and large delays result from numerous outage events. In this respect, MUEs and FUEs have a mutual benefit to cooperate using the proposed approach in order to improve their performance and overcome their limitations. For instance, one of the main benefits of the proposed scheme is the possibility of separating in time uplink transmissions from cooperating MUE and FUE, allowing for cross-tier interference avoidance at the FAP`s side. In other words, the MUE exclusively grants the helping FUE a portion of its superframe to transmit in exchange for cooperation \cite{Osvaldo1}. Establishing a D2D link between the MUE and the FUE comes with a number of advantages, notably due to the low transmit range. In detail, when MUEs and FUEs are close, it is possible to leverage D2D communication at high rates. Clearly, the low transmit range also implies low average transmit power, which allows energy savings at the MUE side. Moreover, a direct link between MUEs and FUEs can also lead to the introduction of novel services and application which require a direct link between MUEs and FUEs.
Therefore, in this paper, we introduce a holistic approach in which we study cross-tier cooperation in a macrocell-femtocell network accounting for delay, power constraints, and optimization of the rewarding mechanism. In summary, our key contributions are the following:

\begin{itemize}
       \item We design a framework for macro-femto cooperation in which the end user benefit is quantified in terms of both throughput and delay.
       \item We tackle the macro-femto coexistence using a cooperative game theoretical approach, by formulating a coalitional game in which MUEs and FUEs are the players. We show that the game is in partition form as it takes into account the external interference between the formed coalitions.
       \item A distributed coalition formation algorithm is proposed through which MUEs and FUEs self-organize to reach the recursive core of the game.
       \item Within each coalition we apply a generalized optimization algorithm so as to maximize the FUE´s revenue, by adequately partitioning the available superframe and setting the transmit power for serving the MUEs in the coalition.
\end{itemize}

The proposed approach enables the MUEs and femtocells to self-organize and jointly establish a D2D link with a FUE, which will access the core network through an FAP access. These operations rely on self-organizing capabilities at the FUEs and MUEs and minimally involve the MBS, since it not notified until the players are actually cooperating. Moreover, the proposed approach is independent of the access policy in use at the FAP side, and could be applied even when the latter adopts a closed-access policy (or when it is congested in open access mode, or its maximum allowable MUEs is reached in hybrid access mode). System level simulations show that the proposed coalition formation algorithm yields significant gains, in terms of average rate per MUE, reaching up to $205\%$ compared to the non-cooperative case, for a network with $N=200$~femtocells.

The rest of this paper is organized as follows. In Section~\ref{sec:sm}, we describe the considered system model and analyze the limitations of the non-cooperative approach. In Section~III we model the macro-femto cooperation as a coalitional game and discuss its properties. In Section~IV, we describe how to optimize the main parameters in the game and provide a distributed algorithm for coalition formation. Simulation results are discussed in Section~V and finally conclusions are drawn in Section~VI.

\section{System Model}\label{sec:sm}

\noindent Consider the \emph{uplink} direction of an Orthogonal Frequency Division Multiple Access (OFDMA) macrocell network (e.g., an LTE-Advanced or WiMAX macrocell) in which $N$ FAPs are deployed. These FAPs are underlaid to the macrocell frequency spectrum, and, within the femtocell tier, neighboring FAPs are allocated over orthogonal frequency subchannels\footnote{We assume that upon startup each femtocell senses the spectrum occupation of the adjacent FAPs and, based on that, it occupies a disjoint set of subchannels, thus, avoiding  interference from the FAPs in proximity \cite{garcia,lp,gustavo,yangg}.}. Let $\mathcal{N}= \left \{ 1,\dots,n,\dots,N \right \}$ and $\mathcal{M}= \left \{ 1,\dots,m,\dots,M \right \}$ denote, respectively, the sets of all FAPs and MUEs in the network. Every FAP $n \in \mathcal{N}$ serves $L_n$ FUEs. Let $\mathcal{L}_n= \left \{ 1,\dots,l,\dots,L_n \right \}$ denote the set of FUEs served by an FAP $n \in \mathcal{N}$.

The packet generation process at each MUE-MBS link is modeled as an M/D/1 queuing system\footnote{Other queue types, e.g., M/G/1 can be considered, without loss of generality.}, in which packets of constant size are generated using a Poisson arrival process with an average arrival rate of $\lambda_m$, in bits/s. Similarly, the link between FUE $l$ and its belonging FAP is modeled as an M/D/1 queuing system with Poisson arrival rate of $\lambda_l$. In the non cooperative approach, the MBS offers MUE $m$ a link transmission capacity (measured in bits/s) of:
\begin{equation}\label{eq:R_m_nc}
\mu_m^{NC} = B \log\Big(1+\frac{\left | H_{m,0} \right |^2 P_m}{\sum_{l\in \Phi_{l}^{0}}\left | H_{l,0} \right |^2 P_l +\sigma^{2}}\Big),
\end{equation}

\noindent where $B$ is the bandwidth of a subchannel, $\left | H_{m,0} \right |^2$ is the channel gain between MUE $m$ and the MBS denoted by subscript $0$, $P_m$ is the power used at MUE $m$, $\Phi_{l}^{0}$ is the set of FUEs operating on the same subchannel as MUE $m$, $\left | H_{l,0} \right |^2$ is the channel gain between FUE $l$ and the MBS, $P_l$ is the power used at FUE $l$ and $\sigma^{2}$ is the noise variance of the symmetric additive white Gaussian noise~(AWGN). The quality of the signal received at the MBS is generally limited by the signal strength, since the MUE-MBS link is often in non line-of-sight~(NLOS) and corrupted by the channel fluctuations and interference from FUEs. In contrast, the femtocell coverage is characterized by higher signal to noise ratio, resulting from the shorter distance between FUE and FAP, and more favorable channel conditions. However, due to the nature of underlay spectrum access, FAPs are limited by the interference from nearby MUEs and by capacity in terms of number of available spectral resources. As a matter of fact, each FAP $n$ provides a generic FUE $l \in \mathcal{L}_n$ with a link transmission capacity of :

\begin{equation}\label{eq:R_i_nc}
\mu_l^{NC} = B \log\Big(1+\frac{\left | H_{l,n} \right |^2 P_l}{\sum_{m\in \Phi_{m}^{n}}\left | H_{m,n} \right |^2 P_m +\sigma^{2}}\Big),
\end{equation}

\noindent where $B$ is the bandwidth of one assigned subchannel, $\left | H_{l,n} \right |^2$ is the channel gain between FUE $l \in \mathcal{L}_n$ and its belonging FAP $n$, $\Phi_{m}^{n} \subset \mathcal{M}$ is the set of MUEs operating on the same subchannel as FUE $l\in \mathcal{L}_n$, $\left | H_{m,n} \right |^2$ is the channel gain between MUE $m$ and FAP $n$. One of the aims of this work is to evaluate the effects of cross-tier interference, thus, the transmission capacity in (\ref{eq:R_i_nc}) only accounts for the interfering MUEs. However, the proposed solution can still be applied with some modifications whether a central or distributed frequency planning is adopted.

The probability of successful transmission can be computed as the probability of maintaining the SINR above a target level $\gamma_m$ and $\gamma_l$, respectively for a MUE or a FUE, and expressed as:

\begin{equation}\label{eq:Ps}
\begin{matrix}
Pt_m=\Pr \left \{\frac{\left | H_{m,0} \right |^2 P_m}{\sum_{l\in \Phi_{l}^{0}}\left | H_{l,0} \right |^2 P_l +\sigma^{2}} \geq \gamma_m\right \},
\\
Pt_l=\Pr \left \{\frac{\left | H_{l,n} \right |^2 P_l}{\sum_{m\in \Phi_{m}^{n}}\left | H_{m,n} \right |^2 P_m +\sigma^{2}} \geq \gamma_l \right \}.
\end{matrix}
\end{equation}

To reduce the outage in MUE-MBS transmissions, a Hybrid Automatic-Repeat-ReQuest protocol with Chase Combining~(HARQ-CC) is employed at the medium access control layer \cite{parkvall}. In this scheme, erroneous packets at the destination are preserved so that they can be soft-combined with retransmitted ones. In general, this procedure, carried out at the MUE side, is highly costly since the MUE has to spend additional power for packet retransmission. Consequently, the effective input traffic $\tilde{\lambda}_m $ from an MUE $m$, accounting for a maximum of $D$ retransmissions is:

\begin{equation}\label{eq:lambda_m}
\tilde{\lambda}_m=\lambda_m \sum_{d=1}^{D}Pt_m(1-Pt_m)^{d-1}.
\end{equation}

\noindent We consider M/D/1 queueing delay for the MUEs $m \in \mathcal{M}$, and thus the average waiting time can be expressed by Little`s law \cite{little} as:
\begin{equation}\label{eq:delay_m}
D_m^{NC}= \frac{\tilde{\lambda}_m}{2\mu_m^{NC}(\mu_m^{NC} - \tilde{\lambda}_m)}.
\end{equation}

\noindent Note that once a transmission on a MUE-MBS link drops due to an outage event, it is reiterated up to $D$ times (otherwise dropped), and the increased congestion represented in (\ref{eq:lambda_m}) produces an average higher delay at the end user, as expressed in (\ref{eq:delay_m}).

\section{Femtocell Cooperation as a Coalitional Game in Partition Form}\label{sec:GF}
\noindent In this section, we formulate the problem of cooperation between FUEs and MUEs as a coalitional game in partition form, whose solution is the concept of the recursive core. The aim of the proposed cooperative approach is to minimize the delay of the MUE transmissions through FUE assisted traffic relay, considering bandwidth exchange as a mechanism of reimbursement for the cooperating FUEs.

In existing wireless networks, FUEs and MUEs are typically scheduled independently, regardless of the access policy used at the FAP side. However, the objectives of the FAPs and the MUEs are intertwined from different viewpoints. At the FAP side, high interference level can be due to MUEs operating over the same subchannel which, consequently limits the achievable rates. At the MUE side, poor signal strength reception may result in a high number of retransmissions and, hence, higher delays.
To overcome this, we propose that upon retransmissions, an MUE might deliver its packets to the core network by means of FUE acting as relay terminal. In this case, at each relay FUE, the incoming packets are stored and transmitted in a First-In First-Out (FIFO) fashion on the access line through its own FAP. We model each relay FUE as a M/D/1 queue and use the Kleinrock independence approximation \cite{KL00}. For the relay FUE, cooperation incurs significant costs in terms of delay and spectral resources, since the FUE relays the combined traffic $\tilde{\lambda}_l$ over its originally assigned subchannels. Therefore, it is reasonable to assume that FUEs will willingly bear the cooperation cost only upon a reimbursement from the serviced MUEs.
We propose that, upon cooperation, the MUE autonomously delegates a fraction $\alpha$ ($0< \alpha \leq 1$) of its own superframe to the serving FUE $l$. At the relay FUE $l$, the portion $\alpha$ is further decomposed into two subslots according to a parameter $0 < \beta_l \leq 1$. The first subslot $\alpha\beta_l$ is dedicated to relay MUE`s traffic. The second subslot of duration $\alpha(1-\beta_l)$ represents a reward for the FUE granted by the serviced MUE, and it is used by the FUE for transmitting its own traffic. This method is known in the literature as spectrum leasing \cite{Osvaldo1} or bandwidth exchange \cite{manrelay} and it represents a natural choice for such kind of incentive mechanisms. The above approach is applied to MUEs with one assigned subchannel, nevertheless, it could be extended to the case of multiple assigned subchannels with some modifications in the negotiating phase. In that case, the relay FUE should be initially informed on the subchannels the MUE can potentially lease. Then, the FUE would communicate its preference, according to the highest gain it can achieve, for a given $\alpha$ and $\beta_l$.

\begin{figure}[!t]
    \centering
      \vspace{+2.0cm} \centerline{\epsfig{figure=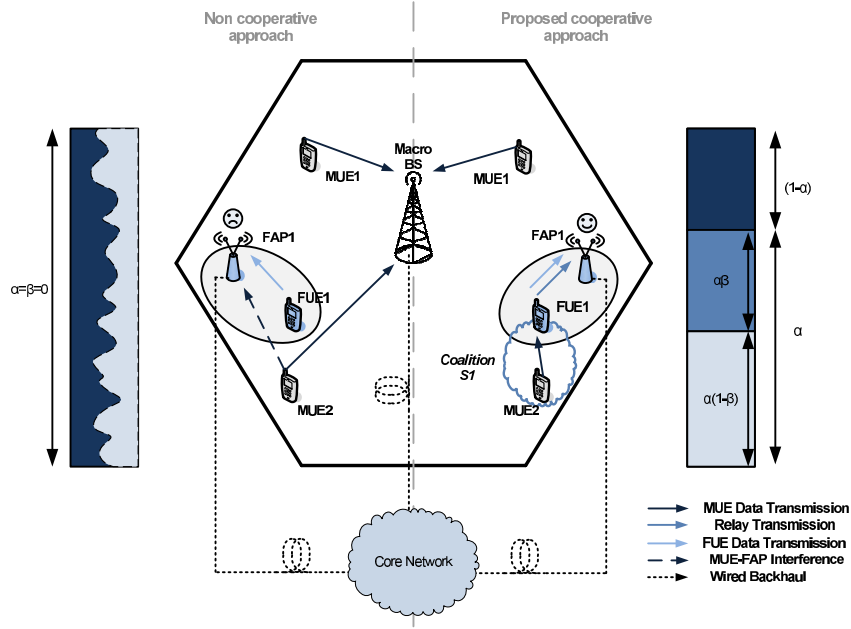, scale=0.6}}
        \vspace{0.8pt}
   \caption{A concept model of the proposed solution compared to the traditional non-cooperative approach.}
   \label{fig:scenario}
\end{figure}

It is worth emphasizing that the proposed solution could still be applied in conjunction with an open, hybrid or closed access policy. Moreover,  MUE transmissions would not require additional spectral resources, as the entire proposed scheme operates without changing the original spectrum allocation in both the femtocell and the macrocell tiers (since it occurs on a D2D link). Figure~\ref{fig:scenario} illustrates the considered scenario compared to the traditional transmission paradigm.

Note that, this concept solution allows to align and separate in time the transmissions allowing to avoid interference at the FAP from the MUEs within the coalition. In order to do that, we assume that operations are synchronized at the system level through IP based synchronization techniques such as IEEE 1588 \cite{synch} in standard or enhanced form.
In order to increase their throughput and reduce MUE-to-FAP interference, the FAPs have an incentive to cooperate and relay the MUEs` traffic. In this respect, FUEs may decide to service a group of MUEs, and thus form a coalition $S_l$ in which transmissions from FUE $l$ and MUEs within the same coalition are separated in time. The proposed cooperation scheme can accommodate any relaying scheme such as decode-and-forward or compress-and-forward schemes. In this work, we use a decode-and-forward relay scheme, assuming that a packet is successfully received if the respective SINR satisfies the conditions in (\ref{eq:Ps}). Finally, the achievable service rates for MUEs and FUEs in the cooperative approach become:

\begin{equation}\label{eq:R_c}
\left\{\begin{matrix}
& \mu_m^{C}(\alpha,\beta_l) &=& \min \{(1-\alpha)\mu_m^{R},\: \alpha\beta_l\mu_l^{R}\},
\\
& \mu_l^{C}(\alpha,\beta_l) &=& \alpha(1-\beta_l)\mu_l^{R},
\end{matrix}\right. \end{equation}

\noindent with,
\begin{equation}\label{eq:R_m_r}
\mu_m^{R} = \log\Big(1+\frac{\left | H_{m,l} \right |^2 P_m}{ \sigma^{2}}\Big),
\end{equation}

\begin{equation}\label{eq:R_i_r}
\mu_l^{R}  = \log\Big(1+\frac{\left | H_{l,n} \right |^2 P_l}{\sum_{m\in \Phi_{m}^{n}\setminus S_l}\left| H_{m,n} \right |^2 P_m +\sigma^{2}}\Big).
\end{equation}

\noindent where $\left|H_{m,l}\right |^2$ denotes the channel gain of the relay link from MUE $m \in S_l$ and FUE $l$.
Note that the factor $(1-\alpha)$ in the first term of (\ref{eq:R_c}) is due to the fraction of superframe occupied by the D2D transmission, while the second factor $\alpha\beta_l$ accounts for the fraction occupied by the forward transmission by the FUE. Due to the fact that MUEs are originally assigned orthogonal subchannels, the first hop of the relay transmission is not affected by interference. Moreover, note that by separating the transmissions from MUE and FUE within the superframe, the FUE forward transmissions are affected only by interference from non cooperative MUEs, outside the coalition. At this point, since the FUE may have to transmit independent packets of its own, the input traffic generation (or the packet arrival at the queue of the FAP) has to account for the packets generated at the FUE and at the MUEs for which the FUEs is relaying. Consequently, the effective traffic $\tilde{\lambda}_{l} $ generated by FUE $l$, accounting for the retransmissions becomes:

\begin{equation}\label{eq:lambda_f}
\tilde{\lambda}_{l}= \big(\lambda_l+\sum_{m \in S_l}\tilde{\lambda}_m\big) \sum_{d=1}^{D} Pt_l(1-Pt_l)^{d-1},
\end{equation}

\noindent where $D$ is the maximum number of retransmissions before the packet is dropped, $Pt_l$ and $\tilde{\lambda}_m$ are computed as in (\ref{eq:Ps}), (\ref{eq:lambda_m}) considering that the SINR this time refers to the FUE-FAP link and using the Kleinrock approximation to combine traffic arrival rates from queues in sequence.

We model every D2D link as a M/D/1 queue system and investigate the average delay incurred per serviced MUE. For a given MUE $m$ served by FUE $l$, we express the average delay as:

\begin{equation}\label{eq:delay_c}
D_m^{R}=\frac{\tilde{\lambda}_m}{2\mu_{m}^{R}(\mu_m^{R} - \tilde{\lambda}_m)}.
\end{equation}

\noindent It is important to underline that, to guarantee the stability of the queues, for any MUE $m$ serviced by a FUE in the network, the following condition must hold:

\begin{equation}\label{eq:cond}
\tilde{\lambda}_m <\mu_{m}^{R}.
\end{equation}
In the event where this condition is violated, the system is considered unstable and the delay is considered as infinite. In this regard, the analysis presented in the remainder of this paper will take into account this condition and its impact on the coalition formation process (as seen later, a coalition where $\tilde{\lambda}_m \geq \mu_{m}^{R}$ will never form).
Having considered this, we now define $\tilde{\lambda}_r = \sum_{m \in S_l}\tilde{\lambda}_m \sum_{d=1}^{D} Pt_l(1-Pt_l)^{d-1}$ and $D_l^{C}= \frac{\tilde{\lambda}_r}{2\mu_{l}^{C}(\mu_{l}^{C}-\tilde{\lambda}_r)}$ as the delay at the FUE for transmitting the traffic of the MUEs' in the coalition. Finally, we can compute the average delay for an MUE as a sum over the MUE-FUE and FUE-FAP hops, as:

\begin{equation}\label{eq:delay_c2}
D_m^{C}= D_m^{R} + D_l^{C}.
\end{equation}

We assume that the relay FUE performs half-duplex operations, i.e., they first receive the MUE`s packets in a transmission window wide $(1-\alpha)$ in the subchannel originally utilized at the MUE. Successively, each FUE forwards the MUE`s packets in the next transmission window wide ($\alpha\beta_l$) in a FIFO policy. We further foresee that, once the packets are forwarded towards the core network, they can be traced back to the original source by means of a small packet header which include the mobile user ID.

\subsection{Coalitional Game Concepts}
Coalitional games involve a set of players, who seek to form cooperative groups, i.e., coalitions, in order to improve their performance or gains. A coalitional game is defined by a set of players, i.e., the decision makers seeking to cooperate and a \emph{coalitional value} (which is either a function or a set of vectors) which quantifies the worth of a coalition in a game, i.e., the overall benefit achieved by the coalition. Classical coalitional problems are typically modeled in the characteristic form, in which the utility of a coalition is not affected by the formation of other distinct coalitions \cite{CF00,Game_theory2,GT00}. In contrast, for coalitional games in \emph{partition form}~\cite{GT00}, the value of any coalition strongly depends on how the players \emph{outside} $S_l$ have organized themselves, i.e., which coalitions they formed. Although coalitional games in partition form are inherently complex to solve, they capture realistic inter-coalition effects that arise in many problems, notably in wireless and communication networks. In this context, finding an optimal solution for games in partition form is a challenging task and is currently a topic of high interest in game theory \cite{CF00,Mac1,WS00,MAC,GT00} (and references therein).

In this section, we mathematically model the problem of macrocell-femtocell cooperation as a coalitional game with the FUEs and MUEs being the players. In particular, having defined $\Psi = \mathcal{M} \cup \: \big\{\bigcup_{n \in \mathcal{N}} \mathcal{L}_n\big\}$ as the set of the players in the proposed game, the rate and the delay achieved by the members of any coalition $S_l \subseteq \Psi$ that forms in the network is affected by the cooperative behavior of the users outside $S_l$, i.e., FUEs and MUEs in $\Psi \setminus S_l$, and thus, we remark the following:
\begin{remark}
The proposed game $(\Psi,U)$ is in partition form.
\end{remark}

This property stems mainly from two reasons. First, under the non cooperative approach, MUEs fully utilize the assigned superframe and, hence, transmit for its whole duration. In consequence, non-cooperative FUEs and MUEs allocated over the same subchannel can collide for the whole transmission duration. In contrast, when an MUE and an FUE belong to the same coalition, the MUE transmits for a fraction $(1-\alpha)$, while the remaining fraction is granted at the FUE in exchange for relaying, hence avoiding collisions between coalitional members. Second, cooperating MUEs transmit over a D2D link which is locally established and has a low transmission range. Therefore, when cooperating, the transmit power at the MUEs are sensibly lower when compared to the non-cooperative scheme and the consequent level of interference suffered at the FAPs outside the coalition is generally lower. As a result, the performance of a coalition depends on the partition of the network $\Pi_{\Psi}$ ($\Pi_{\Psi}$ is a partition of $\Psi$). We will henceforth include this dependence in the definition of the achievable rate in (\ref{eq:R_c}) : $\mu_i^{C}(\alpha,\beta_l,\Pi_{\Psi})$, where $i\in\{m,l\}$ (i.e., MUE or FUE). Given this property, one suitable framework for modeling the macrocell-femtocell cooperation is that of a coalitional game in \emph{partition form} with \emph{non transferable utility} which is defined as follows~\cite{Game_theory2}:

\begin{definition}\label{def:partform}
A coalitional game in \emph{partition form} with \emph{non transferable} utility~(NTU) is defined by a pair $(\Psi,U)$ where $\Psi$ is the set of players (i.e., MUEs and FUEs), and $U$ is a mapping such that for every coalition
$S_l \subseteq \Psi$, $U(S_l,\Pi_{\Psi})$ is a closed convex subset of $\mathbb{R}^{\left | S_l \right |}$ that contains
the payoff vectors that players in $S_l$ can achieve.
\end{definition}

As discussed in the previous section, it is clear that MUEs and FUEs have a strong incentive to cooperate to improve their performance using advanced techniques such as relaying and spectrum leasing. Since MUEs and FUEs exhibit a tradeoff between the achievable throughput and the transmission delay, we use a suitable metric to quantify the benefit of cooperation defined as \emph{power} of the network. Indeed, the power is defined as the ratio of maximum achievable throughput and delay (or a power of the delay)~\cite{KL00,PW01,PW02}. Thus, given a coalition $S_l$, composed by a set of $|S_l|-1$ MUEs and a serving relay FUE $l$, we define a mapping function $U(S_l,\Pi_\Psi)$ as:

\begin{equation}\label{eq:payoff}
U(S_l,\Pi_{\Psi})=\Big\{\boldsymbol{x}\in \mathbb{R}^{\left | S_l \right |} \:|\: x_i(S_l,\Pi_{\Psi})= \frac{\mu_i^{C}(\alpha,\beta_l,\Pi_{\Psi})^{\delta}}{D_{i}^{C\,(1-\delta)}}, \: \forall i \in S_l \Big\},
\end{equation}

\noindent where $\delta \in(0,1)$ is a transmission capacity-delay tradeoff parameter to model the service tolerance to the delay. The set $U(S_l,\Pi_{\Psi})$ is a \emph{singleton set} and, hence, closed and convex. Note that, the player's payoff denoted by $x_i(S_l,\Pi_{\Psi})$ directly refers to a ratio between the achievable throughput and the average delay for player $i$ in coalition $S_l$ and quantifies the benefit of \emph{being a member} of the coalition.
In consequence, the game $(\Psi,U)$ is an NTU game in partition form and, within each coalition, the utility of the players is univocally assigned.

\subsection{Recursive core}\label{sec:Rec}
In order to solve the proposed coalition formation game in partition form, we will use the concept of a \emph{recursive core} as introduced in \cite{K01} and further investigated in \cite{K02,Lazlo2009,Lazlo2}. The recursive core is one of the key solution concepts for coalitional games that have dependence on externalities, i.e., in partition form. Due to the challenging aspect of NTU games in partition form, as discussed in \cite{K02,Lazlo2009,Lazlo2} the recursive core is often defined for games with transferable utility where the benefit of a coalition is captured by a real function rather than a mapping. By exploring the fact that, for the proposed game, as seen in (\ref{eq:payoff}) is a \emph{singleton set}, then we can define an adjunct coalitional game $(\Psi,v)$ in which we use, for any coalition $S_l$, the following function over the real line (i.e., similar to games with transferable utility) which represents the sum of the users' payoffs:
 \begin{equation}\label{eq:value}
v(S_l,\Pi_{\Psi})= \left\{\begin{matrix}
&\sum_{i=1}^{\left | S_l \right |} x_i(S_l,\Pi_{\Psi}) ,& if \: |S_l|>1 \:and\: \alpha>0,  \\
&0 ,& otherwise,
\end{matrix}\right.
\end{equation}
as the \emph{value of the game}. Then, for every coalition achieving (\ref{eq:value}), the individual payoffs of the users are given uniquely by the mapping in (\ref{eq:payoff}). By doing so, we are able to exploit the recursive core as a solution concept for the original game $(\Psi,U)$ by solving the game $(\Psi,v)$ while \emph{restricting} the transfer of payoffs to be according to the unique mapping in (\ref{eq:payoff}).

Further, given two payoff vectors $\boldsymbol{x},\boldsymbol{y} \in\mathbb{R}^{|S_l|}$, we let $\boldsymbol{x}>_{S_l} \boldsymbol{y}$ if $x_i \geq y_i$ for all $i \in S_l$ and for at least one $j \in S_l$ $x_j >y_j$. We also define an \emph{outcome} as couple ($\textbf{x},\Pi_{\Psi}$), where $\textbf{x}$ is a payoff vector resulting from a partition $\Pi_{\Psi}$. Further, let $\Omega(\Psi,v)$ denote a set of all the possible outcomes of $\Psi$.

Essentially, the recursive core is a natural generalization of the well-known \emph{core solution} for games in characteristic form, to games with externalities, i.e., in partition form~\cite[Lemma~10]{K01}. In fact, when applied to a game in characteristic function form, the recursive core coincides with the original characteristic form core~\cite{CF00}. The recursive core is a suitable outcome of a coalition formation process that takes into account externalities across coalitions, which, in the considered game, are represented by effects of mutual interference between coalitions. Before delving into the definition of the recursive core, we need to introduce the concept of a \emph{residual game}:

\begin{definition}
A \emph{residual game} ($\mathcal{R},v$) is a coalitional game in partition form defined on a set of players $\mathcal{R}$, after the players in $\Psi \setminus \mathcal{R}$ have already organized themselves in a certain partition. These players that are outside $\mathcal{R}$ are called \emph{deviators}, while the players in $\mathcal{R}$ are called \emph{residuals}.
\end{definition}

\noindent Consider a coalitional game $(\Psi,v)$ and let $S_l$ be a certain coalition of deviators. Then, let $\mathcal{R}=\Psi \setminus S_l$ denote the set of residual players. The residual game $(\mathcal{R},v)$ is defined as a game in partition form over the set $\mathcal{R}$. Clearly, a residual game is still in partition form and it can be solved as an independent game, regardless of how it was generated as discussed in \cite{K01}. To better present this concept, we will provide an intuitive introduction. For instance, when some deviators reject an existing partition and decide to reorganize themselves into a different partition, their decisions will, in general, affect the payoff of the residual players. As a result, the residual players for a new game that is part of the original game (e.g., the game over the whole set $\Psi$), but with a certain part of the partition (composed by deviators) already fixed. In consequence, one of the main attractive properties of a residual game is its consistency as well as the possibility of dividing any coalitional game in partition form into a number of residual games which, in essence, are easier to solve. In fact, any game in partition form can be seen as a collection of residual games, and each one of those can be solved as if it was the original one. The solution of a residual game is known as \emph{the residual core} which is defined as follows:

\begin{definition}
The \emph{residual core} of a residual game $(\mathcal{R},v)$  is a set of possible game outcomes, i.e., partitions of $\mathcal{R}$ that can be formed.
\end{definition}

One can see that given any coalitional game $(\Psi,v)$, residual games are smaller than the original one and therefore computationally easier to analyze.
Given any coalitional game $(\Psi,v)$, the recursive core solution can be found by recursively playing residual games, which, in fact, yields the following definition as per \cite[Definition~2]{K01}:

\begin{definition}\label{def:Reccore}
The \emph{recursive core} of a coalitional game $(\Psi,v)$ is inductively defined in four main steps:

\begin{enumerate}
  \item \emph{Trivial Partition}. Let ($\Psi,v$) be a coalitional game. The recursive core of a coalitional game where $\Psi\!\!=\!\!\mathcal{M} \cup \mathcal{L}_n$ is composed by the only outcome with the trivial partition composed by the single player $i$: $C(\left \{ i \right \}, v )$ = $(v({i}),i) $.
  \item \emph{Inductive Assumption}. Proceeding recursively, consider a larger network and suppose the recursive core $C(\mathcal{R},v)$ for each game with at most $|\Psi|-1 $ players has been defined. Now, we define the assumption $A(\mathcal{R},v)$ about the game $(\mathcal{R},v)$ as follows: $A(\mathcal{R},v)=C(\mathcal{R},v)$, if $C(\mathcal{R},v) \neq \emptyset $ ; $A(\mathcal{R},v)=\Omega(\mathcal{R},v)$, otherwise.
  \item \emph{Dominance.} An outcome $(\textbf{x},\Pi_{\Psi})$ is \emph{dominated} via a coalition $S_l$ if for at least one \\ $(\textbf{y}_{\Psi\setminus S_l},\Pi_{\Psi \setminus S_l}) \in A(\Psi \setminus S_l,v)$ there exists an outcome $((\textbf{y}_{S_l}, \textbf{y}_{\Psi \setminus S_l}), \Pi_{S_l} \cup \Pi_{\Psi \setminus S_l}) \in \Omega(\Psi,v)$ such that $(\textbf{y}_{S_l}, \textbf{y}_{\Psi \setminus S_l})>_{S_l} \textbf{x}$.
 \item \emph{Core Generation.} The recursive core of a game of $\left | \Psi \right |$ players is the set of undominated outcomes and we denote it by $C(\Psi,v)$.
\end{enumerate}
\end{definition}

Note that, in Definition~\ref{def:Reccore}, the concept of dominance in step 3) inherently captures the fact that the value of a coalition depends on the belonging partition. Hence, it can be expressed in the following way: given a current partition $\Pi_{\Psi}$ and the respective payoff vector $\textbf{x}$, an undominated coalition $S_l$ represents a deviation from $\Pi_{\Psi}$ in such a way that the resulting outcome $((\textbf{y}_{S_l}, \textbf{y}_{\Psi \setminus S_l}), \Pi_{S_l} \cup \Pi_{\Psi \setminus S_l})$ is more rewarding for the players of $S_l$, compared to $\textbf{x}$.

Since a partition uniquely determines the payoffs of all the players in the game, the recursive core can be seen as a set of partitions that allow the players to organize in a way that provides them with the highest payoff. It is important to underline that the recursive core is achieved by verifying relevant properties of \emph{rationality, well-definition and efficiency} as discussed in~\cite{K01}. In detail, with \emph{rationality} it is intended that players never choose an inferior (i.e., dominated) strategy, therefore, they always pursue a profitable strategy. The recursive core is also \emph{well-defined} because when it exists, its solution is unique. Furthermore, \emph{efficiency} is a consequence of the fact that there is no preferred in the set composed by the recursive core, and thus, all the included partitions are equivalent in terms of individual payoff.

Given these properties, once a partition in the recursive core takes place, the players have no incentive to abandon it, because any deviation would be detrimental. As a result, a partition in the recursive core is also \emph{stable} since it is a partition which ensures the highest possible payoff for each one of the players who have no incentive to leave this partition.

Similar to many game theoretic concepts such as the core or the Nash equilibrium, the existence of a recursive core for a coalitional game is a key issue. In \cite{K01}, the author shows that the existence of the recursive core requires \emph{at least one residual core} (and not all of them) to be nonempty. In particular, this means that at least a subset of the players in the network must have defined a preference on how to organize themselves, i.e., how to partition the network. Moreover, an empty residual core reflects a case in which the players of the corresponding residual game do not identify any preferred network partition, or in our proposed cooperation scenario, can equivalently choose between cooperating or not.

Therefore, for the proposed coalitional game, the emptiness of a residual core does not happen and this can be justified as follows. As per Definition~\ref{def:Reccore}, the recursive core is evaluated through a sequence of residual games over subsets of players (i.e., MUEs and FUEs, in our case) in the network. When a given residual core is empty, it is still possible to solve a larger game which contains this as a residual game, in a nested fashion. Hence, the existence of the recursive core is in fact guaranteed as long as one can find at least one residual core that is nonempty. Thus, the recursive core is a solution concept that exists for any game in partition form, unless all the residual cores are empty.

In practice, for the proposed coalitional game, the case in which all residual cores (and, thus, the recursive core) are empty is unlikely to emerge. As a matter of fact, this would represent a network in which any partition of mobile users is \emph{equally likely} to form. In a practical wireless network, this would imply that the MUEs and FUEs are indifferent (i.e., achieve the same payoff) between states in which they are actually suppressing interference and relaying their transmissions (e.g., cooperatively using a D2D link with an FUE) and states in which they are actually suffering from this interference and transmitting to the MBS.


In a nutshell, for the proposed coalitional game, one can use the concept of residual cores in order to find a partition in the recursive core, i.e., a stable and efficient partition, as will be further described in the next section.

\subsection{Distributed implementation of the Recursive core}
\noindent Once a coalition $S_l$ has formed, the FUE $l$ optimizes its own payoff by deciding upon $\beta_l$ and the transmit power.
At the FUE´s side, relaying traffic for a set of MUEs incurs a cost that must be taken into account by the FUE before making any cooperation decision. In this paper, we consider a cost in terms of the transmit power that each FUE spends to transmit for MUEs within the same coalition. Namely, a FUE spends $\beta_l P_l^{(R)}$ to relay MUEs´ traffic and $(1-\beta_l) P_l^{(T)}$ for its own transmissions, while the overall transmit power is limited by $P_{max}$ as:

\begin{align}\label{eq:con}
   \beta_l P_l^{(R)} + (1-\beta_l)P_l^{(T)}<P_{max}.
\end{align}

Leased spectrum and transmit power can be finely tuned in order to maximize the payoff of each member in the coalition.
Accordingly, after a coalition has formed, given a value of $\alpha$, FUE $l \in \mathcal{L}_n$ jointly optimizes the transmit power and the parameter $\beta_l$, by solving the following problem:

\begin{align}\label{eq:opt}
   &\max_{\beta_l, P_{l}} \:\:\: x_i(S_l,\Pi_{\mathcal{N}})\\
   &\text{s.t.} \:\:\: 0<\alpha,\beta_l \leq 1 ; \: \beta_l P_l^{(R)} + (1-\beta_l) P_l^{(T)}<P_{max}.
\end{align}

\noindent Mainly, the FUE is fed back with the estimated aggregated interference from the MUEs $m$ outside the coalition (and included in $\Phi_{m}^{n}$), which can be either measured by its own belonging FAP,  or extimated by considering the MUEs in the proximity \cite{zhangg}.
The problem in (\ref{eq:opt}) can be solved using well known optimization techniques such as those in \cite{BO00}.
\begin{algorithm}[!t]
  \footnotesize
  \flushleft
  \begin{minipage}{\linewidth}
      \caption{{\small Distributed coalition formation algorithm for uplink interference management in two-tier femtocell networks}}
      \label{ALG:recursivecore}
      \begin{algorithmic}
          \STATE{\textbf{Initial State:} The network is partitioned by $\Pi_{\Psi}$ = $\mathcal{M} \cup \mathcal{L}_n$ with non-cooperative MUEs and FUEs.}
          \STATE{\textbf{repeat}}
              \STATE\hspace{\algorithmicindent}{\emph{Phase I - Interferer Discovery}}
              \STATE\hspace{\algorithmicindent}{a) Through RSSI measurements, each FUE detects nearby MUEs, active on the same subchannel, and vice versa.}
              \STATE\hspace{\algorithmicindent}{b) For each of the occupied subchannels, each FUE sorts the interfering MUEs from the stronger to the weaker.}
              \STATE\hspace{\algorithmicindent}{c) Based on the measured RSSIs, each MUE $m$ in $\Psi$ sorts the sensed FUEs from the supposedly closer to the farther.}
              \STATE\hspace{\algorithmicindent}{\emph{Phase II - Coalition Formation}}
              \STATE\hspace{\algorithmicindent}\textbf{for all} mutually interfering MUEs and FUEs in $\Psi$ \textbf{do}
              \STATE\hspace{\algorithmicindent}{a) Each MUE and FUE sequentially engages in pairwise negotiations with the strongest discovered interferer, to identify}
              \STATE\hspace*{2.5em}{potential coalition partners.}
              \STATE\hspace{\algorithmicindent}{b) Each MUE and FUE in $\Psi$ estimates the achieved rate and delay and computes its utility
                         $x_i({S},\Pi_{\Psi})$ as in (\ref{eq:payoff}).}
              \STATE\hspace{\algorithmicindent}{c) FUEs and MUEs engage in a coalition formation which ensures the maximum payoff.}
                \STATE\hspace{\algorithmicindent}\textbf{end for}
             \STATE{\textbf{until}}{ any further growth of the coalition does not result in a payoff enhancement of at least one player \textbf{or} decreases the other coalitional members' payoffs.}
             \STATE{\textbf{Outcome of this phase:} Convergence to a stable partition in the recursive core.}
          \STATE\hspace{\algorithmicindent}{\emph{Phase III - Spectrum Leasing and Cooperative Transmission}}
          \STATE\hspace{\algorithmicindent}{a) Within each coalition, the MUEs notify the serving MBS, and connect to the serving FUE through the D2D operations}
          \STATE\hspace*{2.5em}{described in Section \ref{sec:GF}.}
          \STATE\hspace{\algorithmicindent}{b) Each FUE $l \in S_l$ optimizes its payoff by balancing the transmit power and the transmission window $\beta_l$ by solving}
           \STATE\hspace*{2.5em}{the optimization problem in (\ref{eq:opt}). }
      \end{algorithmic}
  \end{minipage}
\end{algorithm}

To reach a partition in the recursive core, the players in $\Psi$ use Algorithm~\ref{ALG:recursivecore}. This algorithm is composed mainly of three phases: Interferer discovery, recursive core coalition formation, and coalition-level cooperative transmission. Initially, the network is partitioned by $ \left |\Psi\right |$ singleton coalitions (i.e., non-cooperating mobile users).
The MBS periodically requests Received Signal Strength Indicators~(RSSIs) measurements from its MUEs to identify the presence of femtocells which might cooperatively provide higher throughput and lower delays through D2D communication. A similar measurement campaign is carried out at the FUE, as requested by the respective FAP. Successively, for each of the potential coalitional partners, the potential payoffs in (\ref{eq:payoff}) are computed, considering the mechanisms of spectrum leasing captured in (\ref{eq:R_c}). Ultimately, each MUE or FUE sends a request for cooperation to its counterpart which ensures the highest payoff.
If both MUEs and FUEs mutually approve the cooperation request, they form a coalition, set up a D2D connection and the MUE acknowledges its MBS about the established connection. Even during the D2D transmission, the MUEs still maintain a connection to the radio resource control of its original MBS.  Being limited by interference, the most eligible partners for FUEs are dominant interfering MUEs, while, vice versa for a MUE, the higher utilities are granted by FUE in the vicinity or experiencing good channel gains. The recursive core is reached by considering that only the payoff-maximizing coalitions are formed. Clearly, this algorithm is distributed since the FUEs and MUEs can take their individual decisions to join or leave a coalition, while, ultimately reaching a stable partition, i.e., a partition where players have no incentive to leave the belonging coalition. Those stable coalitions are in the recursive core at the end of the second stage of the algorithm. Finally, once the coalitions have formed, the members of each coalition proceed to construct a D2D link and perform the operations described in Section \ref{sec:GF}. As a result, intra-coalition uplink interference at the respective FAPs is suppressed and the MUEs achieve lower delays.

The proposed distributed solution significantly reduces the intrisic complexity of the coalition formation problem as it leverages on the formulation of reduced games among mutual interferers within transmit range, which reduces the search space. Moreover, as per step $(b)$ in Phase I, since the dominant interferers are the most eligible to join the FUE's coalition, they are sorted by descending values of the estimated interference they produce and processed accordingly, which further reduces the number of algorithm iterations.

With regards to the convergence of Algorithm~\ref{ALG:recursivecore}, note that the limitation on the cost for cooperation as per $(15)$ limits the number of potential coalitional partners and, thus, the number of combination that the algorithm has to evaluate. Moreover, by choosing the coalitional partners from an ordered list of interfering MUEs, the resulting FUE's payoff is non decreasing after each iteration of the algorithm. Finally, Algorithm~\ref{ALG:recursivecore} terminates at the first iteration, in which an FUE cannot further increase its payoff without being detrimental for the other coalition partners. Therefore, by cooperatively solving the strongest interference, the FUEs achieve the maximum achievable payoff, and, therefore, have no incentives to break away from the belonging coalitions since it would lead to lower payoffs. Thus, the formed coalitions represent a stable network partition which lies in the recursive core.

The proposed  Algorithm~\ref{ALG:recursivecore} converges to a stable partition which is undominated as per Definition~\ref{def:Reccore}. Although the recursive core might include more than one undominated partition, they are all equivalent, in the sense that they provide the same average player's payoff. Furthermore, due to the concept of dominance since a deviation occurs only towards coalition which guarantees a strictly higher payoff, as per step 3) in Definition~\ref{def:Reccore}, a player MUE or FUE has no incentive to deviate towards equivalent partitions in the recursive core, as they provide equivalent average payoffs.


\section{Simulation Results and Analysis}\label{sec:res}

\begin{table}[!t]
\scriptsize
\caption{System Parameters}
\centering
\setlength{\tabcolsep}{3pt}
\begin{tabular}{|c | c || c| c|}
  \hline
  Macrocell radius & 1Km & Max TX power at MUE and FUE: $P_{max}$  &  20 dBm \\
  Femtocell radius ($r$) & 10-50m & Max number of retransmissions ($D$) & 4 \\
  Carrier frequency & 2.0 GHz & Forbidden drop radius (femto) & 0.2m \\
  Number of FAPs   &  1 - 360  & Total Bandwidth & 100 MHz\\
  Number of FUEs per femtocell  &  1  & Subcarrier Bandwidth $B$ & 180 kHz \\
  Number of MUEs per macrocell  & 1- 500 & Thermal Noise Density &  -174 dBm/Hz \\
  Input traffic macro: $\lambda_m$ (femto: $\lambda_l$)  & 150 Kbps & Path Loss Model [dB] (indoor) & $37+ 30\log_{10}$(d[m]) \\
  Min required SINR at the MBS: $\gamma_m$ (FAP: $\gamma_l$)  & 10 dB (15 dB) & Path Loss Model [dB] (outdoor)& $15.3+ 37.6\log_{10}$(d[m])\\
  FAP antenna gain  &  0 dBi & External wall penetration loss &  12dB\\
  Forbidden drop radius (macro) & 50m & Lognormal shadowing st. deviation  & 10 dB  \\
  Number of antennas at the MBS (FAP) &  1 (1) & Shadowing  correlation between FAPs  & 0  \\
  \hline
  \end{tabular}\label{table:par}
\end{table}

\noindent For system-level simulations, we consider a single hexagonal macrocell with a radius of $1$~Km within which $N$ FAPs are underlaid with $M$ MUEs. Each FAP $n \in \mathcal{N}$ serves $L_n=1$ FUE scheduled over orthogonal subchannel, adopting a closed access policy. We set the maximum transmit power at MUEs and FUEs to $P_{max}=20$~dBm, which includes both the power for the serviced MUE's and its own transmissions as in (\ref{eq:con}). Transmissions are affected by distance dependent path loss shadowing according to the 3GPP specifications \cite{3GPP3}.  For both femto users and MUEs, we assume that power control fully compensates for the path loss. Moreover, a wall loss attenuation of $12$~dB affects MUE-to-FUE transmissions. The considered macrocell has $500$ available subcarriers, each one having a bandwidth of $180$~KHz, and dedicates one OFDMA subchannel to each transmissions. As a matter of fact, assigning multiple subchannels to an MUE would extend the produced interference to more than one FAP, and lead to the formation of overlapping coalitions. However, performing coalition formation with multiple membership yields a combinatorial complexity order due to the need for distributing the capabilities of a user among multiple coalitions. Thus, assigning one subchannel enables the formation of disjoint coalitions and optimizes the tradeoff between benefits from cooperation and the accompanying complexity \cite{CF00,Chali}. Further simulation parameters are included in Table~\ref{table:par}. To leverage channel variations, statistical results are averaged on $10000$ simulation rounds.

\begin{figure}[!t]
    \centering
       \centerline{\epsfig{figure=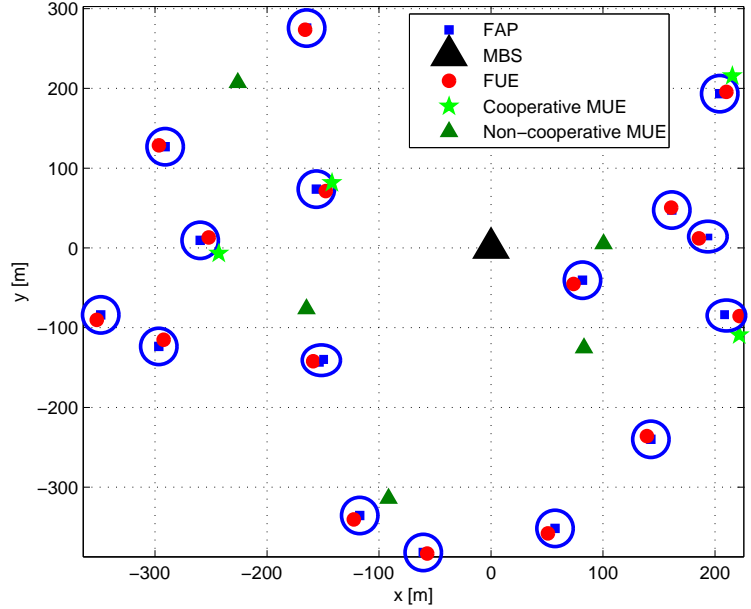, height=90mm}}
        \vspace{-0.4cm}
   \caption{A snapshot of the two-tier femtocell network. The FAPs are modeled by a Poisson point process (squares) and they serve a disc of radius $20$~meters. Triangles represent non cooperating MUEs which communicate with the main base station, represented by a diamond. Stars represent cooperating MUEs which are serviced by the FUE in the coalition (dots).}
   \label{fig:network}
\end{figure}

In Fig.~\ref{fig:network}, we show a snapshot of a femtocell network resulting from the proposed coalition formation algorithm with $N=200$ FAPs that are randomly deployed in the network. The partition in Fig.~\ref{fig:network} lies in the recursive core of the game and is, thus, stable (both FUE and MUE have no incentive to deviate). In this figure, note that although the MUEs are located outside the femtocells, they might be in proximity of a femtocell and potentially interfere with it. If this the case, the FUE has an incentive in forming a coalition with the interfering MUE since it would neutralize its interference. Furthermore, note how the cooperative MUEs are located within the transmission range of a FUE, leveraging on a smaller distance dependent path loss. Conversely, spatially separated MUEs and FUEs are most likely to form singleton coalitions, hence not cooperate. In a nutshell, Fig.~\ref{fig:network} shows how, using the proposed algorithm, the FUEs and the MUEs in a network can self-organize into a partition composed of disjoint and independent coalitions and which is stable, i.e., lies in the recursive core of the game.

\begin{figure}[!t]
    \centering
       \centerline{\epsfig{figure=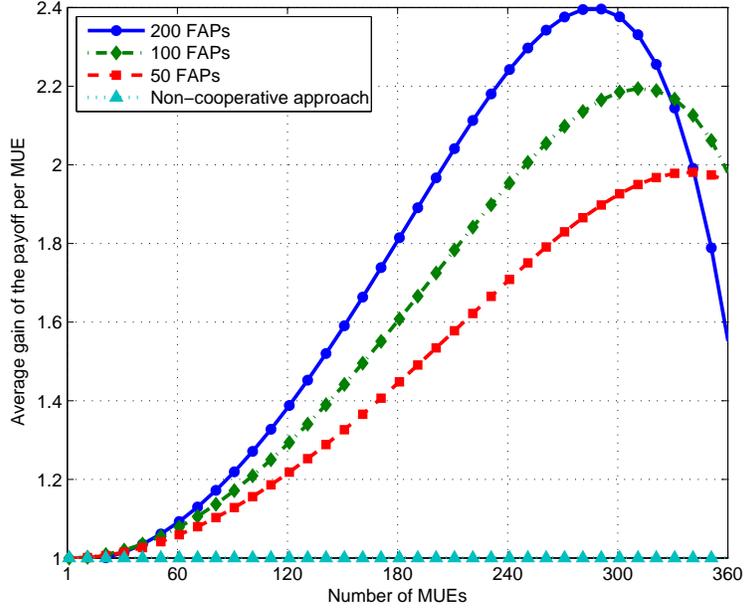, height=90mm}}
        \vspace{-0.4cm}
   \caption{Average gain of individual payoff per MUE, for a network having $N = 50,100,200$ FAPs, $\delta=0.5$, $r=20$m.}
   \label{fig:MUEpayoff}
\end{figure}

In Fig.~\ref{fig:MUEpayoff}, we evaluate the performance of the proposed coalition formation game by showing the average gain of achievable payoff per MUE during the whole transmission time scale as a function of the number of MUEs $M$. We compare the performance of the proposed algorithm to that of the non-cooperative case, for a network with $50,100,200$ FAPs using a closed access policy. The curves are normalized to the performance of the non-cooperative solution. For small network sizes, MUEs do not cooperate with FUEs due to spatial separation. Thus, the proposed algorithm has a performance that is close to the non-cooperative case for $M < 60$. As the number of MUEs grows, the probability of being in proximity of an FUE gradually increases and forming coalitions becomes more desirable. Hence, the MUEs become connected to a nearby FUE which allows for a higher SINR, allowing for high values of payoff. For example, Fig.~\ref{fig:MUEpayoff} shows that cooperating MUE can gain up to $75\%$ with respect than the non-cooperative case in a network with $N=200$ FAPs and $M=160$ MUEs. For larger sizes of the macrocell tier, the coalition formation process eventually saturates and the average gains of cooperation decrease. Further, note that as FUEs in the network not only represent an opportunity of cooperation for the MUEs, but also sources of cross-tier interference, the maximum achievable gains translate towards larger sizes of macrocell tier. In fact, Fig.~\ref{fig:MUEpayoff} clearly shows that the average gain of payoff per MUE increases in the cooperative case as the number of femtocells is large, until each coalition reaches its maximum size. It is also demonstrated that the proposed coalitional game model has a significant advantage over the non-cooperative case, which increases with the probability of having FUEs and MUEs in proximity, and resulting in an improvement of up to $239 \%$ for $M=285$ MUEs.

\begin{figure}[!t]
    \centering
       \centerline{\epsfig{figure=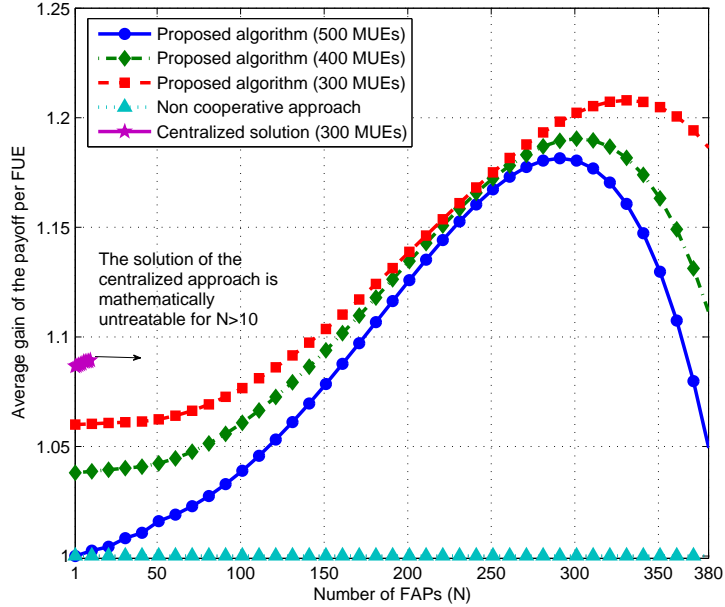, height=90mm}}
        \vspace{-0.5cm}
   \caption{Average gain of individual payoff per FUE, for a network having $M = 300,400,500$ MUEs, $\delta=0.5$, $r=20$m.}
   \label{fig:FUEpayoff}
\end{figure}

In Fig.~\ref{fig:FUEpayoff}, we show the average gain of achievable payoff per FUE as a function of the number of FAPs in the network $N$, for different number of MUEs $M=300, 400, 500$ and normalize the curves to the performance of the non-cooperative solution. As previously seen in Fig.~\ref{fig:MUEpayoff}, cooperation seldom occurs in cases where MUEs and FUEs are spatially separated, as for low numbers of FUEs in the network. Nevertheless, as the density of FAPs increases, coalitions start to take place yielding to higher gains for the FUEs. For instance, Fig.~\ref{fig:FUEpayoff} shows that the average gain of payoff per FUE resulting from the coalition formation can achieve an additional $15\%$ gain with respect to the non-cooperative case, in a network with $N=200$ FAPs and $M=500$ MUEs. However, for larger numbers of FAPs in the network, the average gain in terms of FUE's payoff decreases, as the spectrum becomes more congested and the MUEs in the network have already joined the most rewarding coalitions. Fig.~\ref{fig:FUEpayoff} also shows the comparison with the optimal solution obtained through centralized exhaustive search. For example, Fig.~\ref{fig:FUEpayoff} shows that the performance gap between the centralized and the proposed solution does not exceed $2.6 \%$ for a network of $N=10$ FAPs, while networks with more than $N=10$ FAPs are computationally and mathematically intractable, due to the exponentially increasing number of combinations to be evaluated \cite{CF00}.
Therefore, we demonstrated how cooperation can be beneficial to the FUEs in highly populated areas where the density of interferers (i.e., potential coalitional partners) is high and that the proposed algorithm yields a near optimal performance at a much lower complexity. Finally note that, since the femtocells are orthogonally scheduled, the performance of each FUE in the non cooperative approach is transparent to the density of femtocells in the network.

\begin{figure}[!t]
    \centering
       \centerline{\epsfig{figure=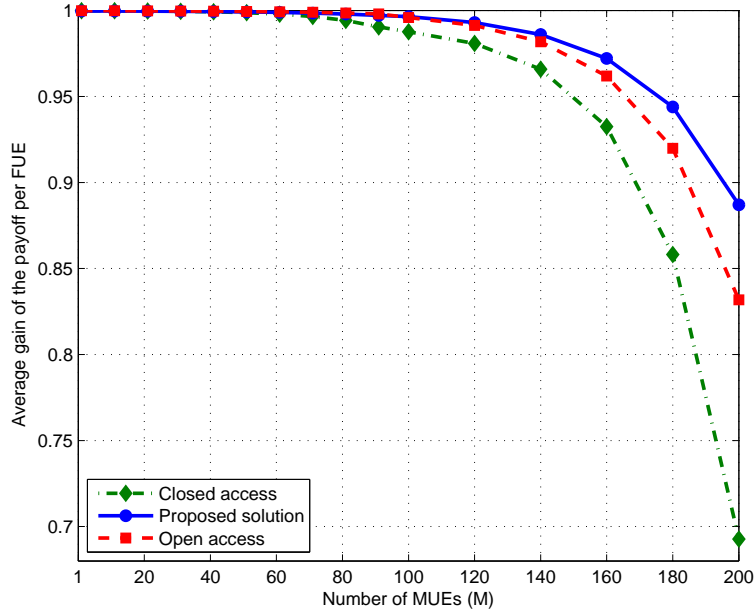, height=90mm}}
        \vspace{-0.4cm}
   \caption{Performance assessment of the proposed network formation algorithm, in terms of average gain of payoff per FUE, for a network having $N=200$ FAPs, under different access policies. $\delta=0.5$, $r=50$m.}
   \label{fig:cross}
\end{figure}

The performance of the proposed coalition formation approach is further assessed in Fig.~\ref{fig:cross}, where we
show the average gain of payoff per FUE as the number of MUEs in the network varies, under different access policies. Here, the curves are normalized to the performance of the closed access policy. Under the open access policy, each FAP has to select a secondary subchannel among the least interfered to schedule the guest MUEs \cite{dlr}. Fig.~\ref{fig:cross} shows that, as $M$ increases, the performance of the FUEs is undermined by the increasing level of interference and a closed access policy may result in a loss of up to $30\%$. An open access policy is more robust to this effect, but it cannot neutralize interfering MUEs which are not in the transmission range. Conversely, our proposed algorithm allows to solve the interference from the dominant neighboring interferers, which are more likely to be in the FUE's transmission range and resulting in a higher gain with respect to the open access policy of $7.6\%$ for a network with $300$ MUEs.

\begin{figure}[!t]
    \centering
       \centerline{\epsfig{figure=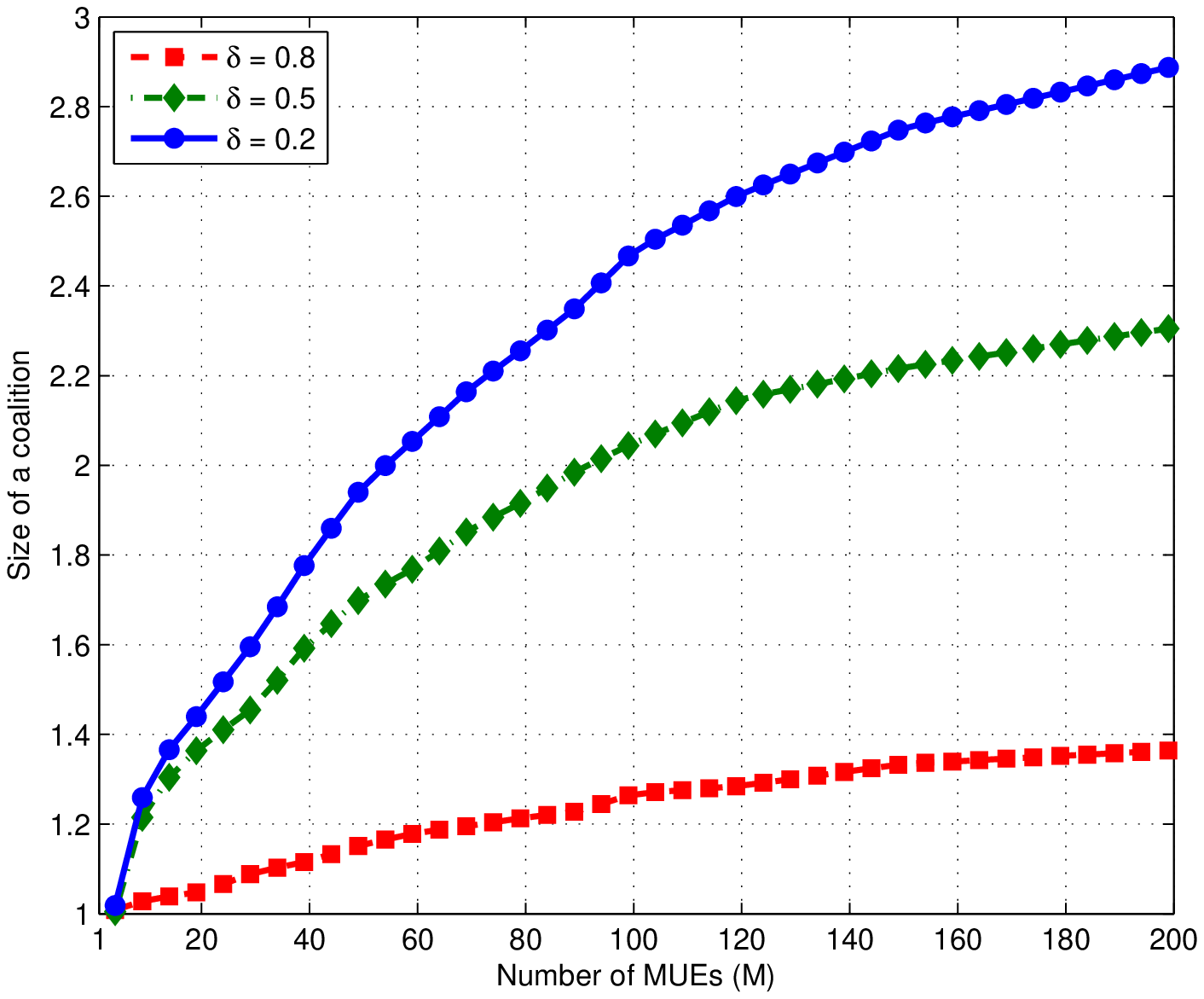, height=90mm}}
        \vspace{-0.4cm}
   \caption{Average coalition size as function of the number of MUEs, for different degrees of delay tolerance, expressed by  $\delta=0.2, 0.5, 0.8$. $N=200$, $r=20$m.}
   \label{fig:coalsize}
\end{figure}

In Fig.~\ref{fig:coalsize}, we show the average size of the coalitions in the recursive core for a QoS parameter $\delta= 0.5$, in a network in which femtocells are extensively deployed ($N=200$). Fig.~\ref{fig:coalsize} shows that due to the high number of cooperation opportunities, the network witnesses an exponential growth of number of MUE-FUE coalitions when the delay constraints are stringent ($\delta=0.2$). For instance, the average coalition size for a network with $M=200$ MUEs is $2.87$. In a less delay constraining case ($\delta= 0.8$), the incentives in cooperation are smaller but still tangible, as demonstrated by a network with $M=200$ MUEs where the average coalition size is $1.39$.
\begin{figure}[!t]
    \centering
       \centerline{\epsfig{figure=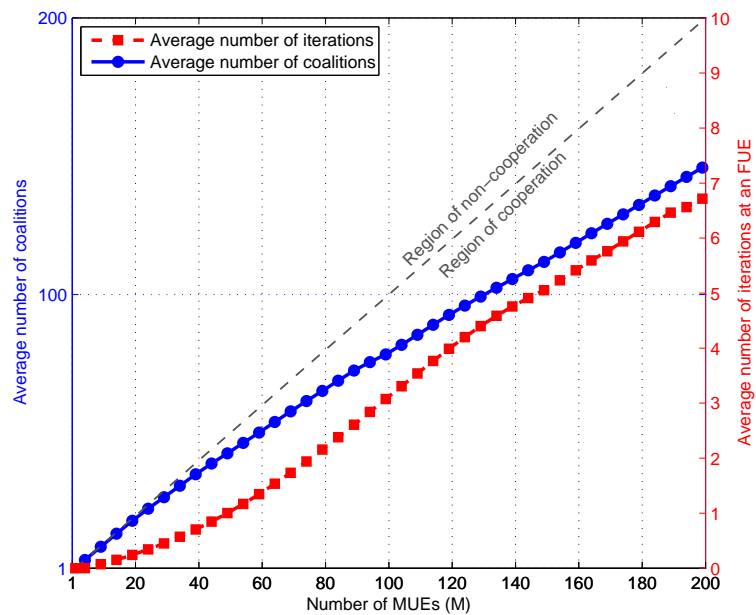, height=90mm}}
        \vspace{-0.5cm}
   \caption{Average number of iterations till convergence and average number of coalitions as a function of the number of MUEs in the network. The bisectrix delimits the area of cooperation and non-cooperation, therefore, the points on the bisectrix represent full non-cooperative MUEs, denoted by singleton coalitions. $N=200$, $\delta=0.5$.}
   \label{fig:numcoal}
\end{figure}
Fig.~\ref{fig:numcoal} shows the growth of the number of coalitions, i.e., the size of a partition in the recursive core, while the number of MUEs increases. Additionally, the average number of iteration in the proposed algorithm is observed. The network is initially organized in a non-cooperative structure where each player (i.e., MUE or FUE) represents a singleton coalition, therefore the number of coalitions equals the number of players in the network (grey dotted line in Fig.~\ref{fig:numcoal}) and, since interferers are out of range of cooperation, the number of iterations is minimum. Initially, for $M<40$ cooperation seldom occurs, due to the large distance between potential coalitional partners. As $M$ increases, the network topology changes with the emergence of new coalitions. For example, when $N=200$ FAPs and $M=200$ MUEs are deployed, $138$ coalitions take place, requiring an average number of algorithm iterations of $6.9$. Therefore, Fig.~\ref{fig:coalsize} and Fig.~\ref{fig:numcoal} show that the incentive towards cooperation becomes significant when the femtocells' spectrum becomes more congested and femtocells are densely deployed in the network. Eventually, for larger $M$, the process of coalition formation is limited by the number of MUEs which a relay FUE can service, given the mechanism of reimbursement in (\ref{eq:R_c}).

\begin{figure}[!t]
    \centering
       \centerline{\psfig{figure=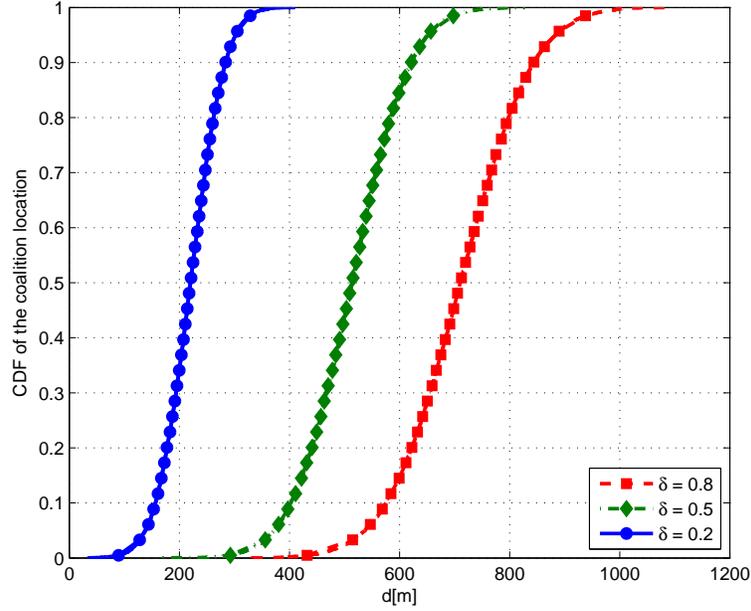, height=90mm}}
        \vspace{-0.5cm}
   \caption{Cumulative distribution function of the distances where the coalitions with more than one user are located from the MBS, for different QoS parameters $\delta=0.2, 0.5, 0.8$.}
   \label{fig:cdf}
\end{figure}

Fig.~\ref{fig:cdf} shows the cumulative distribution function of the distances between the MBS, at the cell center, and the coalitions formed in the network, for $N=200$, $M=200$. This figure shows that the requirement on the QoS, represented by the parameter $\delta$, plays a key role in the coalition formation. In essence, when the delay is more stringent ($\delta =0.2$) than the throughput, as in real time services, cooperation takes place even in the vicinity of the MBS, where higher values of SINR are averagely possible. In contrast, when throughput is more relevant ($\delta=0.8$), coalition formation generally occurs at the cell boundary area, whereas the SINRs at the MBS are limited by the received power. For instance, in Fig.~\ref{fig:cdf} the expected value of the distance from the MBS for a coalition with $\delta =0.2$ is $d=212$ meters, while for $\delta =0.8$ is $d=703$ meters.

\begin{figure}[!t]
    \centering
       \centerline{\psfig{figure=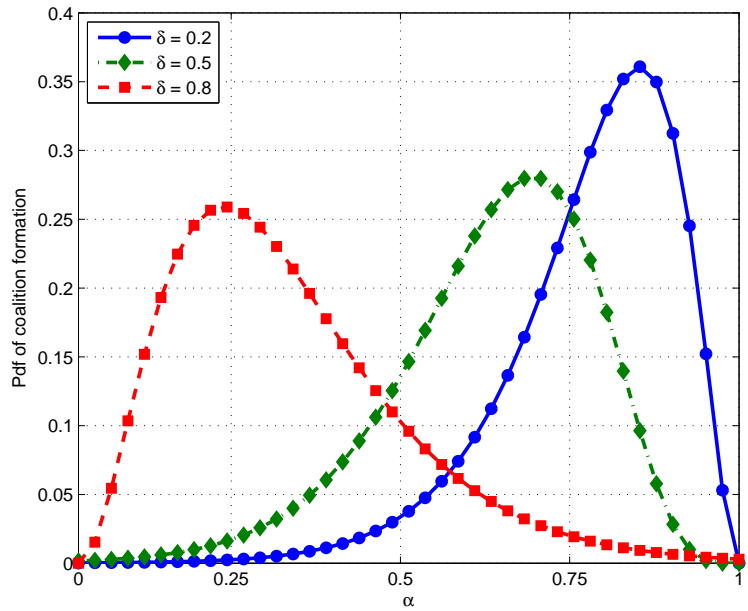, height=90mm}}
        \vspace{-0.5cm}
   \caption{Probability distribution function of coalition formation vs. the superframe fraction $\alpha$ granted to the relay FUE. $M=200$, $N=200$.}
   \label{fig:alpha}
\end{figure}

\begin{figure}[!t]
    \centering
       \centerline{\psfig{figure=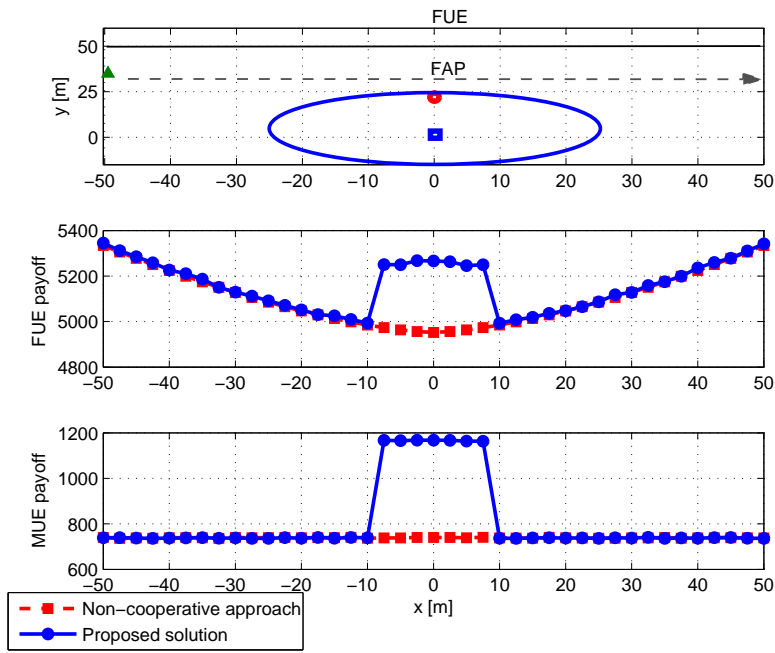, height=90mm}}
        \vspace{-0.5cm}
   \caption{Performance assessment of the proposed network formation algorithm, in terms of average gain of payoff per MUE and FUE at the cell boundary area, in case of a MUE moving on the x-axis in positive direction towards a femtocell. $\delta=0.5$.}
   \label{fig:mobb}
\end{figure}
Fig.~\ref{fig:alpha} shows the probability distribution function of coalition formation as a function of the portion of superframe granted to the relay FUEs, for different $\delta = 0.2, 0.5, 0.8$. This figure demonstrates that, when delay and throughput are equally relevant, an average value of $\alpha= 62\%$ is required by each FUE, for serving an MUE. In contrast, for delay-constrained services, represented by $\delta =0.2$, cooperation becomes more demanding and MUEs have to reimburse the serving FUE for an average value of $\alpha = 78\%$. As a result, we show that the reimbursement mechanism highly depends on the type of service that is required, and the network power is a metric which plays a key role in the coalition formation.

Fig.~\ref{fig:mobb} provides a comparison of the average individual payoff of both cooperative and non-cooperative approaches as a function of the mobility range of a MUE. We consider, from different positions, an MUE close to the macrocell boundary and interfering with a femtocell which adopts a closed access policy. While the MUE is out of the transmission range of the FUE, cooperation cannot be established, thus, the interference from the MUE is unresolved. Conversely, although being located outside of the femtocell and behind a wall, the MUE is serviced by the FUE when the mutual distance is approximately $9.5$ meters, yielding to a significant improvement in terms of respective payoffs. Fig.~\ref{fig:mobb} demonstrates that the proposed solution can lead to an improvement at the MUE side of up to $41\%$. Note that our solution applies not only to the closed access policy, but to all the general cases where an MUE cannot be served by a FAP, although it is harmfully interfering with it (for instance, when the MUE is within interference but out of the FAP transmission range).
\begin{figure}[!t]
    \centering
       \centerline{\psfig{figure=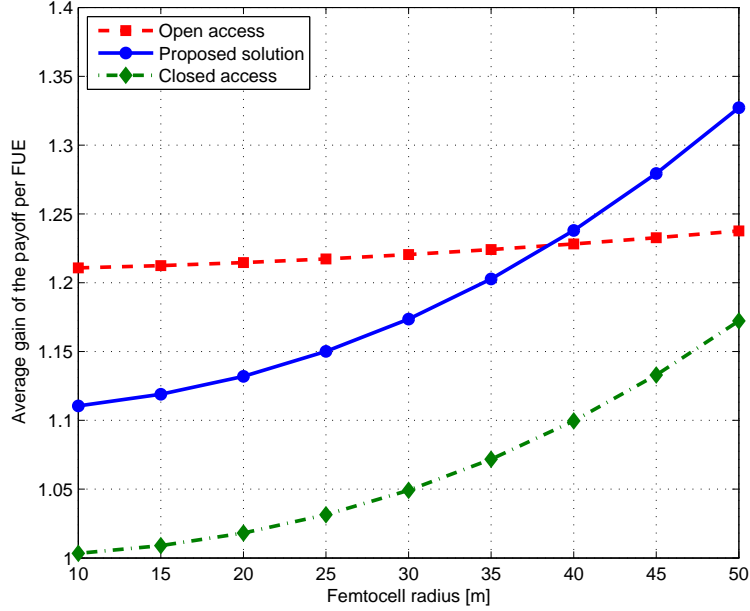, height=90mm}}
        \vspace{-0.5cm}
   \caption{Comparison between the proposed solution and the existing access policies in terms of FUE's payoff vs the size of the femtocell. $M=200$, $N=200$, $\delta=0.5$.}
   \label{fig:radius}
\end{figure}

In Fig.~\ref{fig:radius} we compare our approach to different access policies in terms of average individual FUE's payoff as a function of a femtocell transmission range. The curves are normalized to the performance of the closed access policy for $r=10$ meters. For small femtocell radius, which are currently included in 3GPP specifications\cite{3GPP3}, an open access policy can better protect the FAP from cross-tier interference with respect to a closed access policy. However, when the femtocell radius increases, the FAP is more insulated from the outer interference when located at the cell center. Thus, the closed and open access policies gradually converge. Our proposed solution becomes more beneficial in those cases where, despite the access policy being open, the MUE cannot reach the FAP, leading to a maximum gain of $6\%$ with respect to the open access policy and $14\%$ to the closed access policy, for a femtocell radius of $50$ meters.

\section{Conclusions}\label{sec:conc}

\noindent In this paper, we have introduced a novel framework of cooperation among FUEs and MUEs, which has a great potential for upgrading the performance of both classes of mobile users in next generation wireless femtocell systems.
We formulated a coalitional game among FUEs and MUEs in a network adopting a closed access policy at each femtocell. Further we have introduced a coalitional value function which accounts for the main utilities in a cellular network, namely transmission delay and achievable throughput. To form coalitions, we have proposed a distributed coalition formation algorithm that enables MUEs and FUEs to autonomously decide on whether to cooperate or not, based on the tradeoff between the cooperation gains, in form of increased throughput to delay ratio, and the costs in terms of leased spectrum and transmit power. We have shown that the proposed algorithm reaches a stable partition which lies in the recursive core of the studied game. Results have shown that the performance of MUEs and FUEs are respectively limited by delay and interference, therefore, the proposed cooperative strategy can provide significant gains, when compared to the non-cooperative case as well as to the closed access policy.


\def\baselinestretch{0.80}
\bibliographystyle{IEEEtran}
\bibliography{references}


\end{document}